\documentclass[twocolumn]{aastex631}
\usepackage{xfrac,graphicx,color,float,balance}

\def\ltsima{$\; \buildrel < \over \sim \;$}
\def\simlt{\lower.5ex\hbox{\ltsima}}
\def\gtsima{$\; \buildrel > \over \sim \;$}
\def\simgt{\lower.5ex\hbox{\gtsima}}

\newcommand{\av}{$A_{\rm V}$}
\newcommand{\tdust}{$T_{\rm dust}$}
\newcommand{\jwst}{{\it JWST}}
\defcitealias{kokorev24a}{K24}
\defcitealias{akins24a}{A24}

\newcommand {\um}{$\mu$m}

\newcommand{\msun}{{\rm\,M$_\odot$}}

\newcommand{\lsun}{{\rm\,L$_\odot$}}

\shorttitle{Dust in Little Red Dots}
\shortauthors{Casey et al.}

\begin{document}

\title{Dust in Little Red Dots}

\correspondingauthor{Caitlin M. Casey}
\email{cmcasey.astro@gmail.com}
\author[0000-0002-0930-6466]{Caitlin M. Casey}
\affiliation{The University of Texas at Austin, 2515 Speedway Blvd Stop C1400, Austin, TX 78712, USA}
\affiliation{Cosmic Dawn Center (DAWN), Denmark}

\author[0000-0003-3596-8794]{Hollis B. Akins}\altaffiliation{NSF Graduate Research Fellow}
\affiliation{The University of Texas at Austin, 2515 Speedway Blvd Stop C1400, Austin, TX 78712, USA}

\author[0000-0002-5588-9156]{Vasily Kokorev}
\affiliation{The University of Texas at Austin, 2515 Speedway Blvd Stop C1400, Austin, TX 78712, USA}

\author[0000-0002-6149-8178]{Jed McKinney}\altaffiliation{NASA Hubble Fellow}
\affiliation{The University of Texas at Austin, 2515 Speedway Blvd Stop C1400, Austin, TX 78712, USA}

\author[0000-0003-3881-1397]{Olivia R. Cooper}\altaffiliation{NSF Graduate Research Fellow}
\affiliation{The University of Texas at Austin, 2515 Speedway Blvd Stop C1400, Austin, TX 78712, USA}

\author[0000-0002-7530-8857]{Arianna S. Long}\altaffiliation{NASA Hubble Fellow}
\affiliation{The University of Texas at Austin, 2515 Speedway Blvd Stop C1400, Austin, TX 78712, USA}

\author[0000-0002-3560-8599]{Maximilien Franco}
\affiliation{The University of Texas at Austin, 2515 Speedway Blvd Stop C1400, Austin, TX 78712, USA}

\author[0000-0003-0415-0121]{Sinclaire M. Manning}\altaffiliation{NASA Hubble Fellow}
\affil{Department of Astronomy, University of Massachusetts Amherst, 710 N Pleasant Street, Amherst, MA 01003, USA}


\begin{abstract}
\jwst\ has revealed a ubiquitous population of ``little red dots''
(LRDs) at $z\simgt4$, selected via their red rest-frame optical
emission and compact morphologies.  They are thought to be reddened by dust,
whether in tori of active galactic nuclei or the interstellar medium
(ISM), though none have direct dust detections to date.
Informed by the average characteristics of 675 LRDs drawn from the
literature, we provide ballpark constraints on the dust
characteristics of the LRD population and estimate they have average
dust masses of $\langle M_{\rm
  dust}\rangle=(1.6^{+4.8}_{-0.9})\times10^{4}$\,\msun, luminosities
of $\langle L_{\rm IR}\rangle=(8^{+3}_{-5})\times10^{10}$\,\lsun\ and
temperatures of $\langle T_{\rm dust}\rangle=110^{+21}_{-36}$\,K.
Notably, the spectral energy distributions are thought to peak at
$\sim$100\,K (rest-frame 20-30\,\um) regardless of heating mechanism,
whether AGN or star formation. LRDs' compact sizes
  $R_{\rm eff}\sim100\,pc$ are the dominant factor contributing to
  their low estimated dust masses.
Our predictions likely mean LRDs have, on average, submillimeter
emission a factor of $\sim$100$\times$ fainter than current ALMA
limits provide.  The star-to-dust ratio is a factor $\sim$100$\times$
larger than expected from dust formation models if one assumes the
rest-optical light is dominated by stars; this suggests stars do not
dominate.  Despite their high apparent volume density, LRDs contribute
negligibly (0.1\%) to the cosmic dust budget at $z\gtrsim4$ due to
their low dust masses.
\end{abstract}

\keywords{}

\section{Introduction}\label{sec:intro}

A key discovery of \jwst's first few years has been the remarkable
ubiquity of a population of $z\gtrsim4$ compact ``little red dots''
\citep[LRDs;][]{labbe23a,matthee24a}.  The population is characterized
by compact (spatially unresolved) morphologies and red rest-frame
optical colors.  Many also have a faint, blue component in the
rest-frame ultraviolet (UV), resulting in a characteristic `V-shaped'
SED\footnote{Some works explicitly select for the V-shaped SED as
photometric redshifts are much better constrained with a detected
rest-frame UV component
\citep{labbe23a,labbe23b,kocevski24a,kokorev24a}.}.  The population of
LRDs at $z\gtrsim4$ is surprisingly common, with volume densities
$\gtrsim$10$^{-5}$\,Mpc$^{-3}$, making up a few percent of the galaxy
population at these epochs.  And yet, a similar population at low-$z$
-- both compact and red -- seem not to exist or are orders of
magnitude more rare \citep{kocevski23a}.

The first spectra of the population
\citep{greene23a,kocevski23a,matthee24a} revealed that nearly all LRDs
($\sim$80\%) have broad Balmer lines (H$\alpha$ and H$\beta$ with
FWHM\,$\gtrsim$\,2000\,km\,s$^{-1}$) consistent with a direct view of
the broad-line region of a luminous active galactic nuclei (AGN).
Such broad-line systems are typically seen alongside blue rest-frame
UV continuum emission from the accretion disk
\citep{antonucci93a,peterson06a,hickox18a} as both are spatially
coincident (the broad-line region is on order light-days and
immediately adjacent to the accretion disk on a.u. scales); where one
is seen, so is the other.  The juxtaposition of the red rest-frame
optical colors of LRDs combined with broad lines marks a fundamental
puzzle for our understanding of the physics of LRDs; one implies
significant dust attenuation ($A_{\rm V}\approx2-4$) while the other
seems inconsistent with such attenuation.

The presence of significant dust attenuation has a few important
ramifications.  First, the intrinsic luminosities of LRDs may be
larger than observed; high intrinsic luminosities at high redshifts
then increase potential tension with fundamental models of the
formation of the first massive black holes and/or the galaxies around
them.  It then becomes important to understand the dominant source of
the continuum emission in LRDs: is it from a reddened accretion disk,
or reddened starlight from the host galaxy?  And where is the relevant
dust -- in a hot torus on $\sim$pc scales, or in the ISM extended on
$\sim$100\,pc scales?

Answering these questions about LRDs are core to understanding why
their volume densities are so high.  If LRDs are dominated by AGN
emission -- both across all wavelengths and the full population --
then they imply a tremendous overabundance of supermassive black holes
(SMBHs), about two orders of magnitude \citep{kokorev24a,kocevski24a}
larger than was expected from pre-JWST observations of the bolometric
quasar luminosity function \citep{shen20a}.  Alternatively, if
starlight is contributing significantly to the continuum
\citep{wang24b}, then several LRDs may push the limits of stellar mass
assembly in $\Lambda$CDM (fundamentally limited by the cosmic baryon
fraction times the halo mass, $M_\star\le f_{\rm b}M_{\rm halo}$, with
very high stellar baryon fractions $\epsilon_\star\sim0.3-0.5$).

While the broad lines point to ubiquitous
AGN\footnote{c.f. \citet{baggen24a} argues that stellar
kinematics could lead to broadened Balmer lines given the high stellar
densities \citep{hopkins10a,misgeld11a}.}, there is evidence the
continuum may not be from AGN.  For example, a majority ($>$95\%) of
LRDs lack dominant hot dust emission ($\sim$500-2000\,K) from a torus
which may be expected at luminosities similar to the bolometric
luminosity of the accretion disk itself\ \citep{hickox18a}.  It is
suggested that such hot torus dust should be directly visible using
\jwst's MIRI instrument if the reddening in LRDs is due to torus dust.
Current \jwst/MIRI constraints of LRDs at $>$10\,\um\ imaging largely
show that LRDs have flat spectra in F$_\nu$ through the mid-infrared,
consistent with no torus emission \citep{williams23a,wang24a}.

If the dust responsible for reddening the rest-optical spectra of LRDs
is not in the form of $>$1000\,K torus dust, then it must be present
on scales of the interstellar medium (ISM) instead, at significantly
colder temperatures, perhaps as low as 30\,K .  Here we use the
measured characteristics of LRDs to place boundary conditions on dust
in LRDs.  \S~\ref{sec:samples} presents the LRD samples we use, and
\S~\ref{sec:calc} uses their characteristics to infer dust masses,
luminosities, and temperatures.  \S~\ref{sec:fullsed} presents
predictions for the rest-UV through radio broadband SED of LRDs,
\S~\ref{sec:context} measures their dust mass density contribution,
and \S~\ref{sec:discuss} concludes. 

\section{Reference Samples of LRDs}\label{sec:samples}

\begin{figure*}
\centering
  \includegraphics[width=0.49\textwidth]{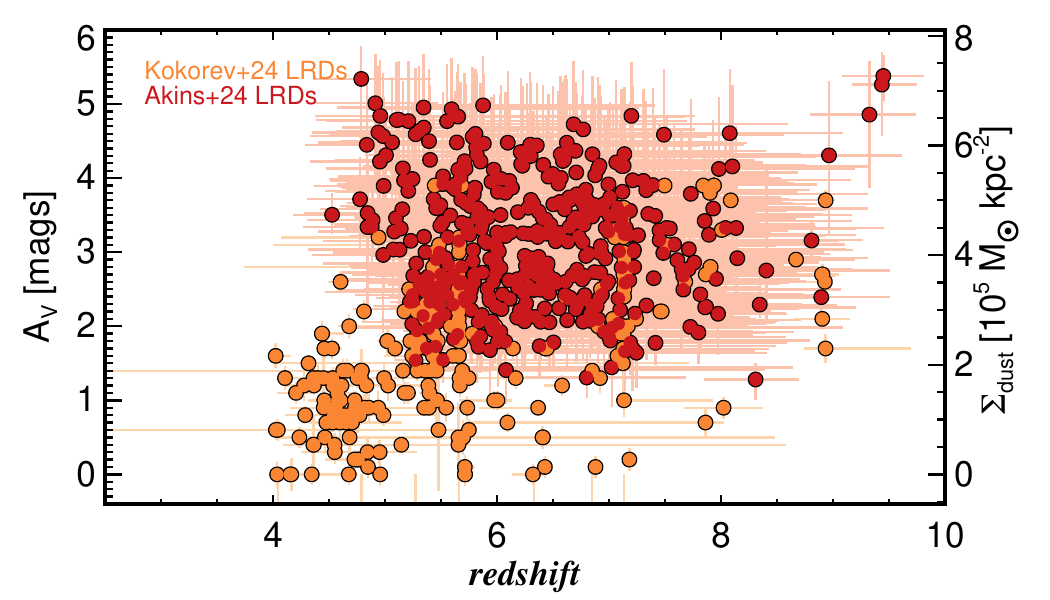}
  \includegraphics[width=0.49\textwidth]{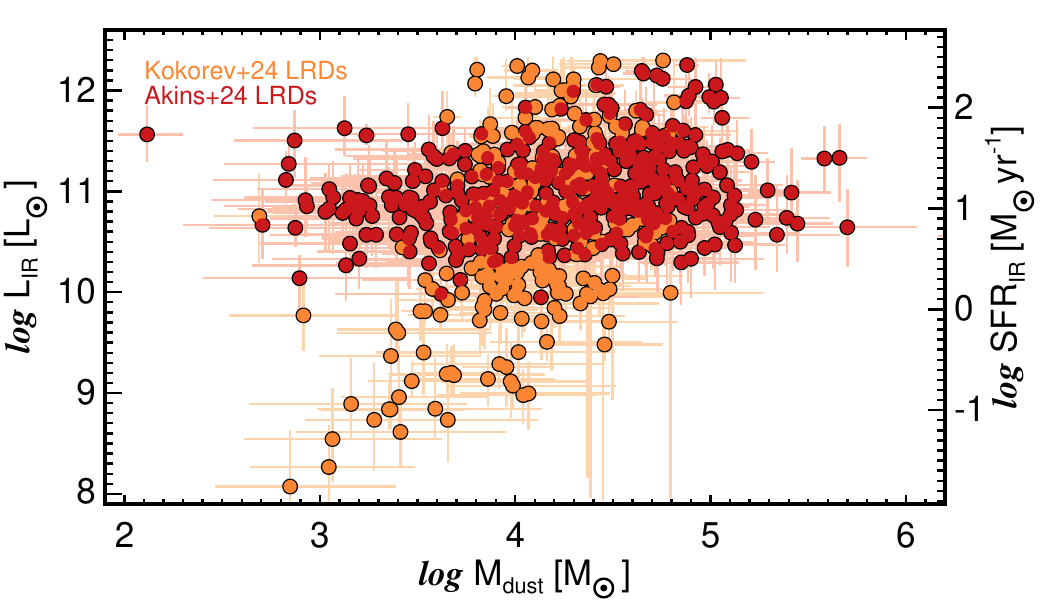}\\
  \includegraphics[width=0.49\textwidth]{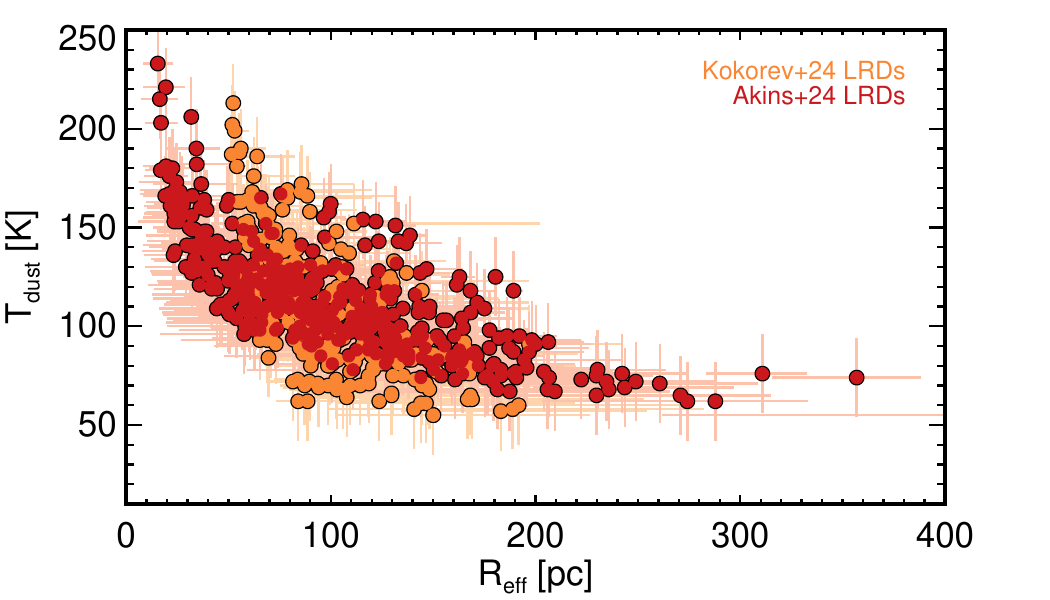}
\caption{{\it Top left:} Photometric redshift vs. visual magnitudes of
  extinction (\av) inferred through SED fitting presented in
  \citetalias{kokorev24a} and \citetalias{akins24a}.  On the right
  y-axis we show the mapping from \av\ to dust mass surface density
  using Eq.~\ref{eq:1}.  {\it Top right:} The 
  estimated dust masses (M$_{\rm dust}$) and dust
  luminosities (L$_{\rm IR}$) for the sample.
  $L_{\rm IR}$ is agnostic to the source of emission (whether AGN or
  star formation) while $M_{\rm dust}$ depends on size, thus may be
  regarded as an upper limit.  {\it Bottom:} Reported size
  measurements for LRDs against the interpolated peak SED dust
  temperature, which is derived from the combination of M$_{\rm dust}$
  and $L_{\rm IR}$.}
\label{fig:properties}
\end{figure*}

We use two literature samples of little red dots in this work.  The
first is the LRD sample described in \citet{kokorev24a} (hereafter \citetalias{kokorev24a}), selected
across $\sim$640\,arcmin$^2$ of \jwst\ key extragalactic surveys
including CEERS \citep[\#\,1345;][]{bagley22a}, PRIMER in COSMOS and
UDS (\#\,1837), JADES \citep[\#1180, 1210, 1286,
  1287;][]{eisenstein23a,eisenstein23b} and FRESCO
\citep[\#\,1895;][]{oesch23a} in GOODS-S and JEMS
\citep[\#\,1963;][]{williams23b}.  \citetalias{kokorev24a} describe the
sample selection in detail in their \S~3.1; to summarize, it consists
of a red color cut at $\lambda>2$\,\um\ (with either ${\rm
  F200W-F356W>1.0}$ or ${\rm F277W-F444W>0.7}$, depending on targeted
redshift) and a moderately blue color cut between
$1$\um$<\lambda<2$\,\um\ (with either ${\rm F115W-F150W<0.8}$ or ${\rm
  F150W-F200W<0.8}$).  Sources are also selected to be spatially
unresolved in F444W.  This results in selection of 260 LRDs brighter
than 27.7\,AB in F444W.

The second LRD sample we use is from \citet{akins24a} (hereafter
\citetalias{akins24a}) selected over 0.50\,deg$^2$ of the COSMOS-Web
Survey \citep[\#1727;][]{casey23a}.  They present 434 LRDs brighter
than 27.5\,AB in F444W.  The \citetalias{akins24a} selection is more
extreme (requiring ${\rm F277W-F444W>1.5}$) and there is no explicit
selection for a blue rest-frame UV slope (at
$1$\,\um\,$\lesssim\lambda\lesssim2$\,\um).

These two works are chosen here because of their large sample size,
together consisting of 675 independently identified LRDs.  Key
differences between the samples are their redshift and color
distributions; the \citetalias{kokorev24a} sample has a broader
redshift distribution spanning $3<z<9$ while the \citetalias{akins24a}
sample is restricted to $5<z<9$.  This difference is largely due to
the different color cuts (the red threshold is more extreme and
filters more restrictive for the \citetalias{akins24a} sample).
Figure~\ref{fig:properties} shows redshift vs. \av\ derived in both
works, highlighting that the LRDs in \citetalias{akins24a} tend to be
a bit redder, more highly attenuated and higher redshift.

Another key difference is the approach to spectral energy distribution
(SED) fitting in either work.  \citetalias{kokorev24a} presumes all
light (both continuum and emission lines) is dominated by an AGN, and
therefore SED fitting is done using quasar templates.
\citetalias{akins24a} uses both a set of galaxy templates and quasar
templates to decompose the SEDs.  We find that the magnitudes of
extinction in the visual band, \av, is consistent whether derived
through galaxy or quasar templates. In other words, our results are
agnostic to the SED methodology used to measure $z_{\rm phot}$, \av,
or the rest-frame absolute UV magnitude $M_{\rm UV}$ (calculated using
\av, $z_{\rm phot}$, and the photometry).

Despite their unresolved morphologies, the sizes of LRDs can be
measured when detected at high signal-to-noise.  Both
\citetalias{kokorev24a} and \citetalias{akins24a} present size measurements in the
F444W filter by assuming Sersic (extended) light profiles.
The inferred sizes are largely consistent, with mean sizes $\langle
R_{\rm eff}\rangle=91^{+22}_{-32}$\,pc and $\langle R_{\rm
  eff}\rangle=94^{+33}_{-72}$\,pc, respectively (both smaller than the
PSF).  These measured sizes should be taken with a grain of salt;
though they could be thought of as direct evidence of ISM-scale
emission, Sersic model fits to a control sample of stars results in
statistically similar sizes.  Additional uncertainty on
  the sizes may come from uncertainty in the PSF as well as the
  applicability of a Sersic model. Nevertheless, we adopt the
measured sizes at face value and will discuss how smaller
(or larger) sizes may impact results.

The derived properties we use to constrain the dust characteristics of LRDs are:
\begin{enumerate}
\vspace{-2mm}
\item \av, which maps directly to a dust mass surface density,
  $\Sigma_{\rm dust}$, in the optically-thin regime,
\vspace{-2mm}
\item $R_{\rm eff}$, the half-light radius at F444W, used to map from
  dust mass surface density to total dust mass and also from dust mass
  and luminosity to dust temperature,
\vspace{-2mm}
\item $M_{\rm UV}$, the rest-frame UV absolute magnitude, which can be
  used in conjunction with \av\ to infer the re-radiated dust
  luminosity, and
\vspace{-2mm}
\item SFR and M$_{\star}$ (measured only for the the
  \citetalias{akins24a} sample), which are used as a semi-independent
  check on the derived dust luminosity.
\vspace{-2mm}
\end{enumerate}
\S~\ref{sec:calc} presents the relevant scalings used to derive
 ballpark dust properties.  Despite the differences in
  selection between the two LRD samples, the dust properties we
 estimate for either sample are similar.

\section{Derivation of Dust Properties}\label{sec:calc}

Our goal is to place broad constraints on the dust properties of LRDs
as a population.  No LRDs have yet been detected via their dust
emission in the millimeter, so here we derive their dust properties
from observed \jwst\ photometry in a back-of-the-envelope
sense. Note that our derivation of dust mass, luminosity
  and temperature is agnostic to AGN or stellar explanations for the
  LRD population. An exception to this is if the intrinsic sizes of
  LRDs are significantly smaller than the measured $R_{\rm
    eff}\sim100$\,pc, which would only {\it decrease} the dust mass
  and {\it increase} the dust temperature.  The other exception is in
  the translation of dust luminosity and temperature into spectral
  energy distributions, the details of which differ based on the
  assumed energy source and are presented in \S~\ref{sec:fullsed}.

\subsection{Dust Masses}\label{sec:mdust}

Absolute magnitudes of attenuation maps directly to dust (and gas)
column density.  We use the observed ratio of magnitudes of visual
extinction to hydrogen column density, $A_{\rm V}/N_{\rm
  H}=5.34\times10^{-22}$\,mag\,cm$^2$/H, consistent with measurements
of X-ray observations of H column density and empirical optical
extinction in the Milky Way \citep{bohlin78a,diplas94a,guver09a}.
\citet{draine07b} present a range of dust-to-hydrogen mass ratios (in
their Table~3); we adopt the canonical $M_{\rm dust}/M_{\rm H}=0.01$
(where $M_{\rm H}=M(${\sc H\,i}$+H_{2})$ and the dust-to-gas ratio,
DGR, is equal to ${\rm DGR=M_{\rm dust}/M_{\rm H} \times 1/1.36}$ to
account for the mass of helium).  This should hold for an ISM enriched
to $Z_\odot>0.2$ \citep{remy-ruyer14a}.  This gives
\begin{equation}
A_{\rm V} = 0.74 \bigg(\frac{\Sigma_{\rm dust}}{\rm 10^{5}\,M_\odot\,kpc^{-2}}\bigg)\, {\rm mag}
\label{eq:1}
\end{equation}
as given in \citet{Draine14a} (see also \citealt{aniano12a}).
This corresponds to a uniform, foreground dust screen.
  We proceed with the scaling in Eq~\ref{eq:1} but note that work on
  nearby galaxies \citep{kreckel13a} and distant obscured galaxies
  \citep{hodge24a} advocate for higher dust mass surface densities
  which they fit empirically on physical scales $>$300\,pc, finding
  $A_{\rm V}=0.18 \Sigma_{\rm dust}/10^{5}\,M_\odot\,kpc^{-2}$; this
  would result in $\Sigma_{\rm dust}$ a factor of $\sim$4$\times$
  higher than we quote herein.

To derive a total dust mass, we multiply by an effective
surface area $2\pi R_{\rm eff}^2$, found by integrating the $n=1$
S\'{e}rsic function in polar coordinates from $r=0-\infty$:
\begin{equation}
M_{\rm dust} = 2\pi R_{\rm eff}^2 \Sigma_{\rm dust}
\label{eq:2}
\end{equation}
 The factor of two partially makes up for the
      simplicity of the foreground dust screen model by assuming
      double the dust mass (i.e. a screen in front and behind the
      illuminating source); the screen's density is assumed to fall off
      exponentially in the radial direction.  These dust masses
      may be a factor of $\sim$2 lower than what is presumed using the
      \citet{kreckel13a} $A_{V}-\Sigma_{\rm dust}$ scaling, but within
      the same order of magnitude.

Ideally, one would have direct measurements of the effective surface
area of dust emission, which may be different than the size of the
system as inferred at 4.4\,\um.  Thus an implicit assumption is that
the dust is absorbing light on the same physical scales of the ISM as
measured via starlight in F444W.  

 We wish to emphasize that dust mass is most sensitive to
  the size measurement, $R_{\rm eff}$, and not the $A_{\rm
    V}-\Sigma_{\rm dust}$ scaling. Because LRDs are inferred to be so
  compact ($<$300\,pc), the dust masses do not exceed
  10$^{5}$\,M$_\odot$.  There could, in principle, be a significant
  offset between measured near-IR size and the dust continuum size, as
  one might expect for highly-obscured star forming galaxies like
  DSFGs where optically-thick dust permeates the ISM, leaving only
  small channels of rest-frame UV/optical light to leak out where
  modest attenuation may be inferred \citep{howell10a,casey14b}.
  However, we note that recent {\it JWST} imaging results of a large
  sample of $\sim$300 DSFGs \citep[][see also
    \citealt{hodge24a}]{mckinney24a} show that the population are
  large $>$1\,kpc systems with extended disks and/or disturbed
  morphologies; all are detected in [F444W] brighter than 26.5\,AB
  magnitudes and have extended morphologies distinct from LRDs.

  As an independent test of our derivation of dust masses using
  $A_{\rm V}$ and size, we apply the above method 
 with the popultion of DSFGs from \citet{mckinney24a}
  that have dust detections, comparing dust masses derived from the
submillimeter with those derived from NIRCam. We found consistent results within
$\pm$0.6\,dex, even though the dynamic range of dust
  masses of DSFGs is quite different than our estimates for LRDs.

Figure~\ref{fig:properties} show the derived dust masses of the LRD
  sample, with mean dust mass $\langle M_{\rm
  dust}\rangle=(1.6^{+4.8}_{-0.9})\times10^{4}$\,\msun.

\subsection{Dust Luminosities}

Next we infer the bulk dust luminosities of the sample.  This
corresponds to the light that has been attenuated in the rest-frame
UV/optical and re-radiated at long wavelengths, commonly measured
between rest-frame 8--1000\,\um\ as $L_{\rm IR}$.
In the absence of direct long wavelength constraints, energy balance
implies that the difference between the observed and intrinsic UV
luminosity is what should be absorbed by dust and re-radiated at long
wavelengths as $L_{\rm IR}$.  How much light is absorbed relates to
\av.

In practice, $L_{\rm IR}$ may be found using the source's observed
absolute UV magnitude $M_{\rm UV}$:
\begin{equation}
  L_{\rm IR} (L_\odot)= (10^{0.4(4.83-M_{\rm UV})})(10^{0.4A_{\rm UV}}-1)
\end{equation}
where $A_{\rm UV}$ is the absolute magnitudes of attenuation at
rest-frame UV magnitudes and $L_{\rm IR}$ given in \lsun.  $M_{\rm
  UV}$ is the intrinsic emission expected in the UV from the {\it
  reddened} component of the SED.  This differs from the faint blue
emission that may be present in the rest-frame UV at much lower
observed luminosities.  The relationship between $A_{\rm UV}$ and
\av\ is a function of the attenuation curve; for Milky Way, LMC, and
Calzetti dust attenuation curves, $A_{\rm UV}/A_{\rm V}\approx2.5$
\citep{salim20a}.

Note that we have made some simplified assumptions to get
order-of-magnitude luminosities.  First, we assume that the bolometric
luminosity, in the absence of dust, is well approximated by the
systems' UV luminosities.  Also, as written, L$_{\rm IR}$ above
corresponds to {\it all} re-radiated dust emission, not just emission
from 8--1000\,\um, so a bolometric correction is needed.  However,
this correction factor is $\le$1.2 for all dust temperatures $<$200\,K
and $\le$2 at $<$300\,K; this temperature range is appropriate for
LRDs as we will soon find.

Another way of estimating $L_{\rm IR}$ comes from SED measurements of
star-formation rates and how much of that SFR is obscured by
dust. Note that this is not entirely independent of other SED-derived
characteristics like $M_{\rm UV}$ and \av, but a slightly different
set of assumptions serves as a worthwhile order-of-magnitude
check. \citet{whitaker17a} presents measurements of the obscured
fraction of star formation, $f_{\rm obscured}\equiv {\rm SFR_{\rm
    IR}}/{\rm (SFR_{\rm UV}+SFR_{\rm IR})}$ as a function of stellar
mass, $M_\star$ for galaxies from $0<z<2.5$.  They find no redshift
evolution in $f_{\rm obs}(M_\star)$.  Specifically,
\citet{whitaker17a} find:
\begin{equation}
f_{\rm obscured} = (1+a\exp(b\log_{10}(M_\star/M_\odot)))^{-1}
\end{equation}
where $a=(1.96\pm0.14)\times10^9$ and $b=-2.277\pm0.007$.  An
SED-derived SFR can then be multiplied by $f_{\rm obscured}$ to obtain
${\rm SFR}_{\rm IR}$; then using $\log(L_{\rm IR}/[L_\odot])=\log({\rm
  SFR}/[M_\odot\,yr^{-1}])-9.83$ \citep{kennicutt12a} we get $L_{\rm
  IR}$.  We note that the discrepancy between $L_{\rm IR}$ estimated
via $M_{\rm UV}$ and \av\ and that estimated via ${\rm SFR}$ and
$M_\star$ has a scatter of $\sim$1\,dex and systematic offset of
$\sim$0.4\,dex (such that the latter method estimates higher $L_{\rm
  IR}$ than the former).  This is not entirely surprising given the
known scatter about the \citeauthor{whitaker17a} relation and its
possible evolution towards higher redshift
\citep[e.g. see][]{zimmerman24a}. We estimate a mean IR
  luminosity for the LRD sample of $\langle L_{\rm
    IR}\rangle=(7.9^{+2.9}_{-4.7})\times10^{10}$\,\lsun.

We emphasize that our ballpark estimates could be wildly incorrect for
individual sources, but it should capture the bulk order-of-magnitude
dust characteristics of the population as a whole. An
  extra note of caution comes, again, from highly obscured
  star-forming galaxies like DSFGs; it is not uncommon that energy
  balance can underestimate $L_{\rm IR}$, sometimes by up to an order
  of magnitude \citep{swinbank04a,da-cunha15a,casey17a} when a clumpy
  ISM and high column densities lead to a `decoupling' of the OIR and
  dust SEDs. However, if such an underestimation were to apply for
  LRDs, it would be challenging to hide the additional luminosity; it
  would emerge either in the mid-IR (e.g. MIRI) if contained in a
  compact volume, or in the submillimeter (e.g. ALMA) if spatially
  extended.  Existing upper limits in both regimes suggest such an
  underestimate does not apply to LRDs.

\subsection{Dust Temperatures}\label{sec:temp}

We derive infrared SEDs (thus luminosity-weighted dust temperatures)
by combining our newly-derived constraints on $M_{\rm dust}$ and
$L_{\rm IR}$.  These SEDs are also neutral to
  interpretation as AGN or star-formation as they are simple modified
  blackbodies. Specifically, we use a simple optically-thin modified
blackbody with modest mid-infrared powerlaw component ($\alpha_{\rm
  IR}=4$), emissivity spectral index $\beta=2$ following the
methodology described in \citet{casey12a} and subsequently
\citet{drew22a}.  The dust mass is proportional to the flux density on
the Rayleigh-Jeans side of the blackbody as well as the dust
temperature, while the IR luminosity is proportional to the integral
of the SED.  These specific SED assumptions lead to a relation between
dust temperature, IR luminosity and dust mass of
\begin{equation}
T_{\rm dust} (K) = c \bigg(\frac{L_{\rm IR}}{L_\odot}\times\frac{M_\odot}{M_{\rm dust}}\bigg)^{d}
\end{equation}
where $c=8.1\pm0.5$ and $d=0.169\pm0.004$. The average
luminosity-weighted dust temperature for the full sample is $\langle
T_{\rm dust}\rangle=110^{+21}_{-36}$\,K.

There are many caveats regarding the relationship of dust temperature
to peak wavelength, whether it is mass-weighted or luminosity-weighted
or an optically-thin or thick dust medium
\citep{scoville16a,casey14a}; however, the dust mass column densities
we estimate here are only moderate, which allows for such simplified
assumptions.  By altering our set of assumptions on $\alpha_{\rm IR}$,
$\beta$, and by propagating uncertainties from $M_{\rm dust}$ and
$L_{\rm IR}$, the characteristic uncertainty on \tdust\ is $\sim$12\,K.

One could, in principle, infer a dust temperature using the
Stefan-Boltzmann Law; they are a factor of $\sim$2$\times$ discrepant
from those we find in this work.  An assumed condition of
Stefan-Boltzmann is that the emergent luminosity has passed through an
optically thick medium, which is likely not a valid presumption in the case
of these systems; \citet{burnham21a} presents a detailed discussion of
 applying Stefan-Boltzmann to dust emission in
galaxies (see their \S~5.1 and \S~5.2.1).

\section{The dust SED of LRDs}\label{sec:fullsed}

\subsection{SED Construction}

\begin{figure*}
  \centering
  \includegraphics[width=0.95\textwidth]{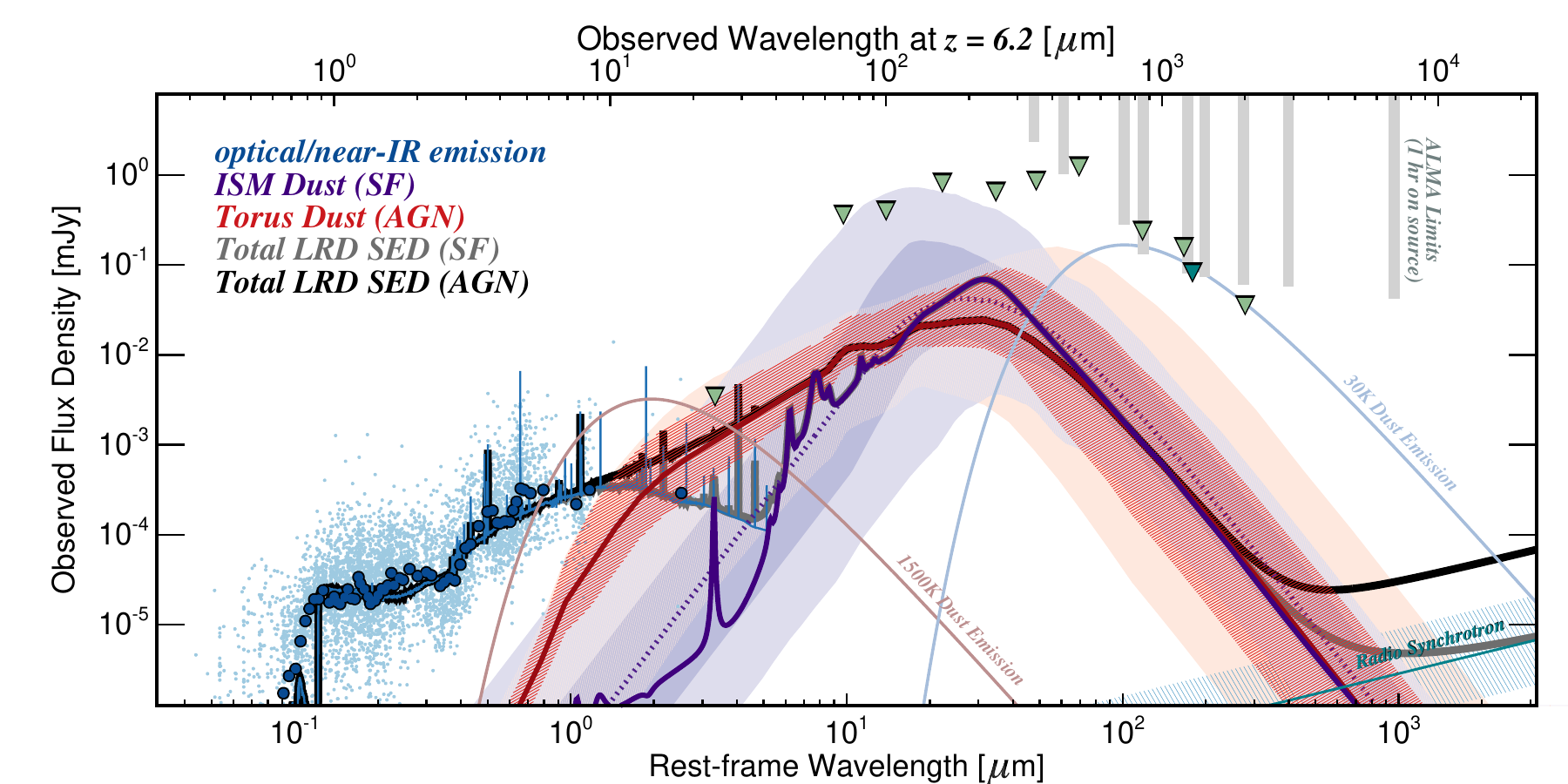}
  \caption{The average rest-frame UV through radio SED of the 675 LRDs
    in this work.  The raw optical/near-infrared photometry of all
    LRDs is shown in light blue points, the median stack in dark blue
    points, and best-fit {\sc Bagpipes} model to the median stack
    (thin-blue line). The purple shaded region denotes the 68$^{th}$
    and 95$^{th}$ percentile confidence regions on the derived dust
    SED discussed in \S~\ref{sec:calc}, and the best-matched
    $\sim$100\,K \citet{draine07a} model is shown in dark purple.  The
    matching radio synchrotron component, assuming the FIR/radio
    correlation holds to high-$z$ \citep{smolcic17a} is shown in teal
    with $\pm$0.5\,dex.  Expectation for AGN torus emission is shown
    in orange, matched to the same predicted $L_{\rm IR}$ as the SED
    in \S~\ref{sec:calc}.  The total SED, summing optical/near-IR,
    dust and synchrotron components, is shown in gray (using the
    \citeauthor{draine07a} star-formation template) and black (using
    the \citeauthor{nenkova08a} torus template); both are made
    available.  Measured 5$\sigma$ limits in the mid/far-IR are shown
    in light green triangles from \citetalias{akins24a} and from
    \citet{labbe23b} in a dark green.  Vertical gray bars show the
    5$\sigma$ continuum depth achievable by ALMA in each of its
    observing bands with 1\,hr on-source integration time (as of ALMA
    Cycle 11 in 2024).}
  \label{fig:sed}
\end{figure*}

We construct an average SED for LRDs from the rest-frame UV through
the radio\footnote{Both average SEDs are made available here:
https://zenodo.org/records/13770535.\nocite{dustlrds}} using the combined
aggregate LRD optical/near-infrared (OIR) photometry from
$HST$/\jwst\ provided in \citetalias{kokorev24a} and
\citetalias{akins24a} and combine this with the dust SEDs derived in
\S~\ref{sec:calc}.  The OIR SED is constructed by converting
photometry to the rest-frame using the best-fit photometric redshift
for each LRD and using the isophotal wavelength of each filter to
convert to effective wavelength of the photometric measurement;
non-detections are included.  Median stacks are calculated in bins
with 100 photometric constraints below $\lambda_{\rm rest}<8000\,{\rm
  \AA}$ and in two additional bins (with $\sim$20 measurements each)
at longer wavelengths.  The median OIR stack, converted back to
observed flux density at the median sample redshift of $\langle
z\rangle=6.2$, is shown in dark-blue points on
Figure~\ref{fig:sed}\footnote{Note this stack is very similar to the
image-plane stack derived in \citetalias{akins24a} for LRDs, which is
reassuring.}.
We fit the stacked photometry using {\sc Bagpipes} \citep{carnall18a}
with the same methodology as \citetalias{akins24a}. Assuming no AGN contribution
toward the rest-UV/optical SED, we infer an average stellar mass of
$\langle M_\star \rangle=(8.1^{+3.1}_{-2.7})\times10^9$\,\msun\ and
attenuation $\langle A_{V}\rangle=1.84\pm0.15$.  We return to the
issue of stellar mass in \S~\ref{sec:stars}.

After generating individual, simple IR SEDs for all LRDs
based on their inferred $M_{\rm dust}$ and $L_{\rm IR}$ 
  in \S~\ref{sec:temp}, we compute the median, 68$^{th}$ and
95$^{th}$ inner confidence intervals, shown in purple on
Figure~\ref{fig:sed}.  Our IR SED modeling is agnostic
  to heating mechanism--star formation or AGN--and thus does not
include complex features from PAH and silicates.

  In the case of a star-forming origin, we also
overplot the best-matched realistic dust template to our median SED
from \citet{draine07a}, which has a Milky Way extinction curve with
0.5\%\ of dust from PAH molecules and a uniform incident radiation
field with $U=10^{5}$, i.e. $10^{5}$ more intense than in the solar
neighborhood's surrounding ISM.  Based on the FIR/radio correlation,
we also show the expected contribution of synchrotron emission from
star-formation following extrapolation from \citet{delhaize17a}.

Figure~\ref{fig:sed} shows the stark contrast of this average warm SED
to a simpler, cold ($\sim$30\,K) modified blackbody oft used to model
the dust component of star formation in the absence of direct FIR
constraints \citep{pope08a,swinbank10a,swinbank14a,casey20a}.  Note
that LRDs' SEDs are substantially warmer thanks to their compact
sizes. Such low dust temperatures would only be feasible at lower dust
surface densities (lower \av) or more spatially extended emission.
While the latter remains a possibility ---implying that the OIR
emission would be dwarfed by a huge dust reservoir---
  such a system would be consistent with $\sim$kpc-scale DSFGs.  As
  discussed in \S~\ref{sec:mdust}, all such DSFGs are spatially-resolved at
  high SNR in F444W. Put another way, it should
be quite surprising that {\it none}
of these 600$+$ LRDs are detected as submillimeter galaxies
\citep[e.g.][]{simpson19a}.

We also explore the relative expected emission from AGN tori
independent of our simple derived SEDs from
  S~\ref{sec:temp}; the expected sublimation temperature of dust with
mixed carbonaceous and silicate composition is $\sim$1500\,K.  A
1500\,K modified blackbody with luminosity scaled from $L_{\rm bol}$
and \av, assuming the OIR continuum is dominated by an AGN is
inconsistent with existing constraints on mid-IR emission from MIRI;
most LRDs show a flat SED in $F_{\nu}$ extending to rest-frame
$\sim$2-3\,\um. However, adopting a more complex emission model for a
dusty torus from \citet{nenkova08a} dramatically shifts the peak of
the rest-frame SED to long wavelengths and relieves this tension with
data.  The \citeauthor{nenkova08a} SED shown in Figure~\ref{fig:sed}
in red is the median of all $>$10$^6$ models in that work based on
clumpy dust distributions around an AGN accretion disk. Its luminosity
is normalized to the same luminosity expected ($L_{\rm IR}$), just now
attributing its heating to AGN instead of star
formation. The \citeauthor{nenkova08a} models are
  consistent with the AGN torus emission model developed by
  \citet{li24a} for LRDs; \citeauthor{li24a} presume a gray
  attenuation law \citep[due to the destruction of small grains around
    AGN, see][]{gaskell04a} with an extended torus, which has the same
  effect as a assuming a clumpy medium.  In both cases, the SEDs peak
  at intrinsically longer wavelengths and relieve tension with the
  flat LRD SEDs measured with MIRI.

\begin{figure*}
  \centering
  \includegraphics[width=0.9\textwidth]{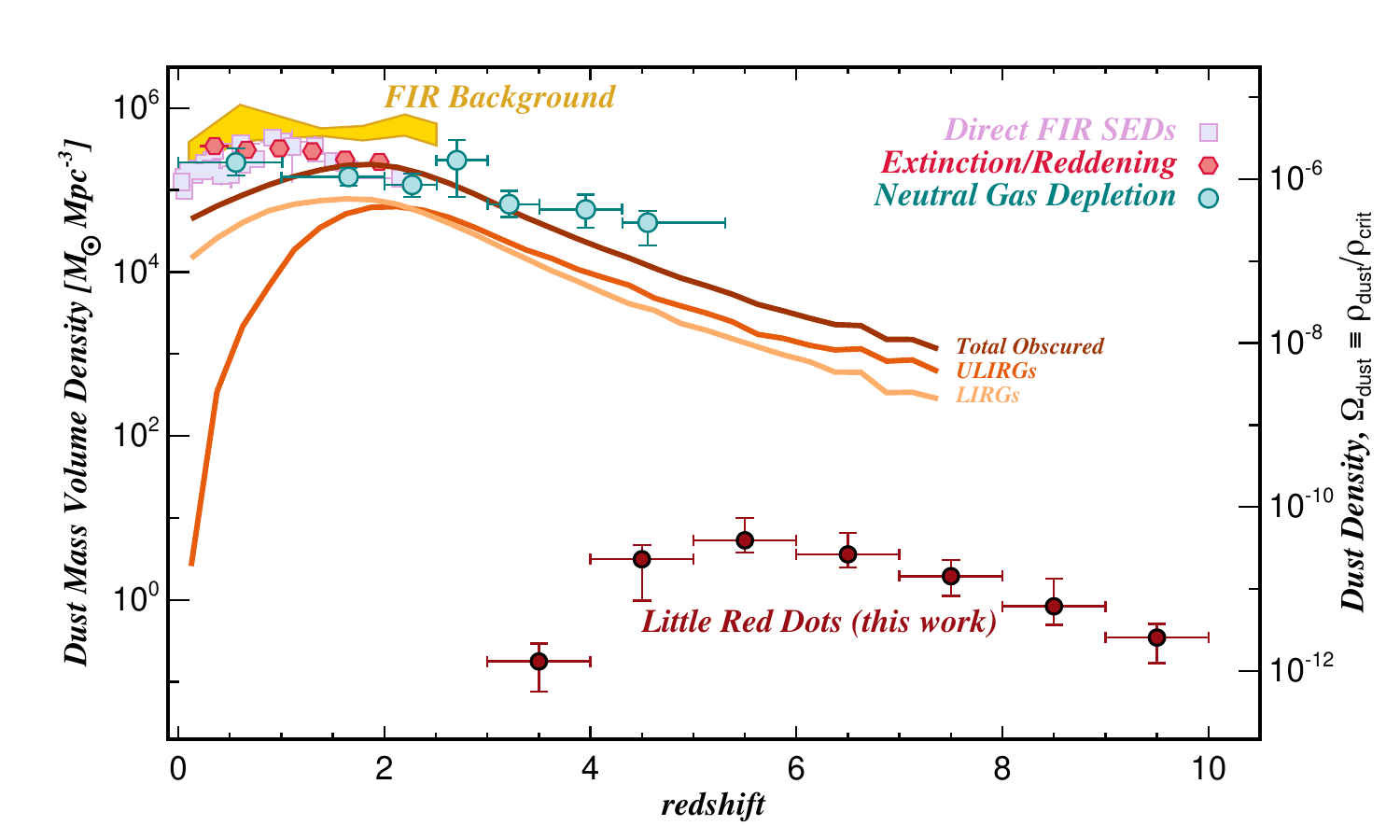}
  \caption{The dust mass volume density of the Universe placing little
    red dots (dark red) in context.  Measurements presented in the
    literature are made from the FIR background directly
    \citep[yellow;][]{thacker13a}, galaxies with directly observable
    FIR SEDs
    \citep[lavender;][]{dunne11a,beeston18a,driver18a,pozzi20a},
    extrapolation from extinction/reddening of OIR SEDs \citep[light
      red;][]{menard10a,menard12a}, and neutral gas depletion
    \citep[teal;][]{peroux20a}.  We overplot the integrated
    contribution from IR-luminous galaxies (with $L_{\rm
      IR}>10^{9}$\,\lsun\ in brown, and specifically from ULIRGs
    (orange; $L_{\rm IR}>10^{12}$\,\lsun) and LIRGs (tan;
    $10^{11}<L_{\rm IR}<10^{12}$\,\lsun) from the \citet{casey18a} model, using
    the measured luminosity function parameters from
    \citet{zavala21a}. Despite their relatively high number density,
    LRDs are thought to only contribute $\sim$0.1\%\ to the cosmic
    dust budget relative to the total, which is dominated by IR
    luminous galaxies.}
\label{fig:dustmassden}
\end{figure*}

The relative consistency of the star-forming dust SED and the AGN
torus dust SED is striking, especially because only the
  star-forming dust SED was derived using our energy balance and size
  measurement presented in \S~\ref{sec:calc}. The \citet{nenkova08a}
models are built to accommodate a complex spatial distribution of dust
clouds distributed in clumps, rather than a uniformly smooth and dense
torus structure on $\lesssim$5\,pc scales \citep[see also discussion
  of obscuration of AGN from][]{maiolino24a}.  As a result, the mean
luminosity-weighted temperature of the aggregate SED is $\sim$100\,K,
which is broadly consistent with what is expected from compact star
formation on $\sim$50--100\,pc scales.  The only significance
deviation between the two SEDs occurs at rest-frame
$\sim$2-10\,\um\ where the relative luminosity of hot dust is higher
for tori.  However, it is clear from Figure~\ref{fig:sed} that
existing constraints in this regime are too few to discriminate
between models when accounting for relevant uncertainties.  Deep MIRI
observations of LRDs will prove crucial to potentially detecting torus
emission, if present.

Note that assuming {\it smaller} sizes for LRDs $\ll$90\,pc would
drive the estimated dust masses down ($M_{\rm dust}\propto R_{\rm
  eff}^2$) and dust temperatures up ($T_{\rm dust}\propto R_{\rm
  eff}^{-2.169}$) such that sizes $\lesssim$25\,pc predict peak dust
temperatures that exceed the sublimation temperature and are in
disagreement with current MIRI constraints.

\subsection{Is the dust detectable with ALMA?}

Figure~\ref{fig:sed} shows depths that can be reached at
5$\sigma$ in each of the ALMA bands (1, 3, 4, 5, 6, 7, 8, 9, and 10)
in an effective 8\,GHz bandwidth with one hour of on-source
integration time\footnote{Calculated using the Cycle 11 version of the
ALMA Observing Tool, Doc. 11.5, version 1.0.}.  Unfortunately these
limits (per source) reach sensitivities two orders of magnitude too
shallow to detect direct dust emission from our predicted models.
This unfortunately leaves direct dust detection via ALMA continuum
observations unlikely for LRDs prior to ALMA's wideband receiver
upgrade (to be completed $\sim$2030), as it would require either a
factor of $>$20 increase in the sample size of LRDs to enable a
stacking experiment on $>$10$^{4}$ sources, or, alternatively, large
investments in time ($\gg$\,days) for very few sources.

\subsection{Reconciling Stellar Mass with Dust Mass}\label{sec:stars}

The stellar mass we derive for the median OIR photometric stack is
{\it high}, nearly 10$^{10}$\,\msun: a factor of $\sim$10$^{5-6}$
larger than our estimated dust masses of LRDs.  Note, of
  course, that this presumes all continuum light is stellar in nature,
  which may not be true. The expected maximum star-to-dust ratio one
expects from supernovae and Asymptotic Giant Branch (AGB) star dust
enrichment models is $\log(M_\star/M_{\rm dust})\approx3-4$
\citep{schneider23b}.  ISM grain-growth would potentially lead to even
lower star-to-dust ratios.  If both mass estimates are accurate, then
it may be likely that optically thick dust (and a patchy ISM geometry)
could lead to the underestimation of the total dust mass.  However,
dust masses that are a factor of $\sim$100$\times$ larger would likely
lead to {\it some} submillimeter detections amongst the sample, for
which there are none (\citetalias{akins24a}).  The other (more likely)
possibility is that stars do not dominate the
  continuum of LRDs.  Put another way, we would expect stellar masses
closer to $\sim10^{7-8}$\,\msun\ by using our dust mass
  estimates and a $\log(M_\star/M_{\rm dust})\approx3-4$; indeed, AGN
light may dominate the rest-optical portion of the SED leading to
systematic overestimation of the stellar mass \citep[see the
  spectroscopic analysis from][]{wang24b}.

\section{Volume Constraints}\label{sec:context}

While the dust reservoirs in LRDs by our estimation are
expected to be rather modest in mass ($\sim$10$^{4-5}$\,\msun), the
relative volume density of LRDs means that they could plausibly
represent the lion's share of dust to have formed at early times,
within the Universe's first $\sim$2\,Gyr. Do they?
To test this hypothesis, we calculate the dust mass volume density
with redshift estimated for the LRDs in this work and compare against
estimates for dusty star-forming galaxies
\citep[DSFGs;][]{blain02a,casey14a}.  To infer the latter, we use the
\citet{casey18a} model framework for building realizations of the
infrared luminosity function and their associated (sub)millimeter
emission.  Specifically, we adopt the measured luminosity function
parameters from \citet{zavala21a} relevant for the $z>2$ Universe,
simulate a 100\,deg$^2$ light cone, then model the dust mass of each
contributor to the IRLF to directly calculate the implied dust mass
volume density.  The associated total dust mass density curves are
shown for all $>10^{9}$\,\lsun\ galaxies, all LIRGs, and all ULIRGs in
Figure~\ref{fig:dustmassden}.  Other measurements of the dust mass
volume density are shown for context, extending out to $z\sim5$ from
\citet{peroux20a}.

Using a total survey area of 0.5\,deg$^2$ for the \citetalias{akins24a} LRD
sample and 640\,arcmin$^2$ for that of \citetalias{kokorev24a}, we compute
the direct contribution of these 675 LRDs to the dust mass volume
density to be on the order of 1--10\,\msun\,Mpc$^{-3}$ (see
Figure~\ref{fig:dustmassden}).  This is 2-3 orders of magnitude lower
than the expected contributions from the more spatially rare DSFG
population at similar epochs.  This indicates that the LRD population,
despite being a relatively common phenomenon, does not contribute
significantly to the cosmic dust budget given that individually their
masses are quite low, $\sim10^{4-5}$\,\msun.  While DSFGs are a much
more dominant contributor
at $z\sim5-7$, it is important to
remember that DSFGs themselves are more rare but have dust reservoirs
that are individually 100-1000$\times$ more massive than estimated for
LRDs.

\section{Summary}\label{sec:discuss}

We have used two large literature samples of little red dots 
to put basic constraints on their dust characteristics.  We use
measured \av, $M_{\rm UV}$, and $R_{\rm eff}$ to
derive dust masses, luminosities and temperatures.  Given their
compact morphologies, their dust masses are estimated to be relatively
small, averaging $\langle M_{\rm dust}\rangle =
(1.6^{+4.8}_{-0.9})\times10^{4}$\,\msun, with luminosities $\langle
L_{\rm IR}\rangle=(8^{+3}_{-5})\times 10^{10}$\,\lsun\ and dust
temperatures $\langle T_{\rm dust}\rangle=110^{+21}_{-36}$\,K.  Such
dust properties would imply that current ALMA observational limits on
LRDs are two orders of magnitude too shallow to render a detection.

While the nature of dust in LRDs is often thought of as a dichotomy --
either embedded in a nuclear torus on $\sim$pc scales and temperatures
$\sim$1500\,K or diffuse on $\sim$100\,pc scales in the ISM with
temperature $\sim$30\,K -- we find that expectation for dust emission
from clumpy torus models \citep{nenkova08a} are fully consistent with
the predicted dust heated via star formation in the ISM, both peaking
at rest-frame $\sim20-40$\,\um.  This convergence of AGN and ISM dust
models is due to LRDs' compact size (driving the ISM dust to be hotter
than `normal' ISM dust) and that complex clumpy torus dust models lead
to cooler dust emission (when weighted by luminosity) than might
nominally be expected.

The star-to-dust mass ratios inferred for LRDs ($\sim10^{5-6}$) are
much higher than expected for typical dust formation scenarios
($\sim10^{3-4}$).  Either AGN contribute significantly to the
rest-optical emission of LRDs \citep{greene23a,wang24a}, resulting in
an overestimation of the stellar mass if fit directly to the
continuum, or dust masses are significantly higher than predicted
here, leaving the possibility that {\it some} may be detectable via
dust continuum with ALMA.

Given the modest predicted dust masses in LRDs, we find they
contribute only negligibly ($\sim$0.1\%) to the cosmic dust budget,
dwarfed by the contribution from DSFGs who are a factor of
$\gtrsim10\times$ more rare but $\gtrsim100-1000\times$ higher dust
mass.

\begin{acknowledgements}
We are grateful to the anonymous referee and to
Justin Spilker, Jorge Zavala, and Seiji Fujimoto for
helpful discussions that improved the manuscript.  CMC thanks the
National Science Foundation for support through grants AST-2009577 and
AST-2307006 and to NASA through grant JWST-GO-01727 awarded by the
Space Telescope Science Institute, which is operated by the
Association of Universities for Research in Astronomy, Inc., under
NASA contract NAS 5-26555.  HBA acknowledges support from the
Harrington Graduate Fellowship at UT Austin, and HBA and ORC thank the
National Science Foundation for support from Graduate Research
Fellowship Program awards.  VK acknowledges support through UT's
Cosmic Frontier Center as an inaugural CFC Postdoctoral Fellow, and
JM, ASL and SMM thank NASA for support through the Hubble Fellowship
Program.

Authors from UT Austin acknowledge that they work at an institution
that sits on indigenous land. The Tonkawa lived in central Texas, and
the Comanche and Apache moved through this area.  We pay our respects
to all the American Indian and Indigenous Peoples and communities who
have been or have become a part of these lands and territories in
Texas.

\end{acknowledgements}


\begin{thebibliography}{}
\expandafter\ifx\csname natexlab\endcsname\relax\def\natexlab#1{#1}\fi
\providecommand{\url}[1]{\href{#1}{#1}}
\providecommand{\dodoi}[1]{doi:~\href{http://doi.org/#1}{\nolinkurl{#1}}}
\providecommand{\doeprint}[1]{\href{http://ascl.net/#1}{\nolinkurl{http://ascl.net/#1}}}
\providecommand{\doarXiv}[1]{\href{https://arxiv.org/abs/#1}{\nolinkurl{https://arxiv.org/abs/#1}}}

\bibitem[{{Akins} {et~al.}(2024){Akins}, {Casey}, {Lambrides}, {Allen},
  {Andika}, {Brinch}, {Champagne}, {Cooper}, {Ding}, {Drakos}, {Faisst},
  {Finkelstein}, {Franco}, {Fujimoto}, {Gentile}, {Gillman}, {Gozaliasl},
  {Harish}, {Hayward}, {Hirschmann}, {Ilbert}, {Kartaltepe}, {Kocevski},
  {Koekemoer}, {Kokorev}, {Liu}, {Long}, {McCracken}, {McKinney}, {Onoue},
  {Paquereau}, {Renzini}, {Rhodes}, {Robertson}, {Shuntov}, {Silverman},
  {Tanaka}, {Toft}, {Trakhtenbrot}, {Valentino}, \& {Zavala}}]{akins24a}
{Akins}, H.~B., {Casey}, C.~M., {Lambrides}, E., {et~al.} 2024, arXiv e-prints,
  arXiv:2406.10341.
\newblock \doarXiv{2406.10341}

\bibitem[{{Aniano} {et~al.}(2012){Aniano}, {Draine}, {Calzetti}, {Dale},
  {Engelbracht}, {Gordon}, {Hunt}, {Kennicutt}, {Krause}, {Leroy}, {Rix},
  {Roussel}, {Sandstrom}, {Sauvage}, {Walter}, {Armus}, {Bolatto}, {Crocker},
  {Donovan Meyer}, {Galametz}, {Helou}, {Hinz}, {Johnson}, {Koda}, {Montiel},
  {Murphy}, {Skibba}, {Smith}, \& {Wolfire}}]{aniano12a}
{Aniano}, G., {Draine}, B.~T., {Calzetti}, D., {et~al.} 2012, \apj, 756, 138,
  \dodoi{10.1088/0004-637X/756/2/138}

\bibitem[{{Antonucci}(1993)}]{antonucci93a}
{Antonucci}, R. 1993, \araa, 31, 473,
  \dodoi{10.1146/annurev.aa.31.090193.002353}

\bibitem[{{Baggen} {et~al.}(2024){Baggen}, {van Dokkum}, {Brammer}, {de
  Graaff}, {Franx}, {Greene}, {Labb{\'e}}, {Leja}, {Maseda}, {Nelson}, {Rix},
  {Wang}, \& {Weibel}}]{baggen24a}
{Baggen}, J. F.~W., {van Dokkum}, P., {Brammer}, G., {et~al.} 2024, arXiv
  e-prints, arXiv:2408.07745, \dodoi{10.48550/arXiv.2408.07745}

\bibitem[{{Bagley} {et~al.}(2022){Bagley}, {Finkelstein}, {Rojas-Ruiz},
  {Diekmann}, {Finkelstein}, {Song}, {Papovich}, {Somerville}, {Baronchelli},
  \& {Dai}}]{bagley22a}
{Bagley}, M.~B., {Finkelstein}, S.~L., {Rojas-Ruiz}, S., {et~al.} 2022, arXiv
  e-prints, arXiv:2205.12980, \dodoi{10.48550/arXiv.2205.12980}

\bibitem[{{Beeston} {et~al.}(2018){Beeston}, {Wright}, {Maddox}, {Gomez},
  {Dunne}, {Driver}, {Robotham}, {Clark}, {Vinsen}, {Takeuchi}, {Popping},
  {Bourne}, {Bremer}, {Phillipps}, {Moffett}, {Baes}, {Bland-Hawthorn},
  {Brough}, {De Vis}, {Eales}, {Holwerda}, {Loveday}, {Liske}, {Smith},
  {Smith}, {Valiante}, {Vlahakis}, \& {Wang}}]{beeston18a}
{Beeston}, R.~A., {Wright}, A.~H., {Maddox}, S., {et~al.} 2018, \mnras, 479,
  1077, \dodoi{10.1093/mnras/sty1460}

\bibitem[{{Blain} {et~al.}(2002){Blain}, {Smail}, {Ivison}, {Kneib}, \&
  {Frayer}}]{blain02a}
{Blain}, A., {Smail}, I., {Ivison}, R., {Kneib}, J., \& {Frayer}, D. 2002,
  \physrep, 369, 111, \dodoi{10.1016/S0370-1573(02)00134-5}

\bibitem[{{Bohlin} {et~al.}(1978){Bohlin}, {Savage}, \& {Drake}}]{bohlin78a}
{Bohlin}, R., {Savage}, B., \& {Drake}, J. 1978, \apj, 224, 132,
  \dodoi{10.1086/156357}

\bibitem[{{Burnham} {et~al.}(2021){Burnham}, {Casey}, {Zavala}, {Manning},
  {Spilker}, {Chapman}, {Chen}, {Cooray}, {Sanders}, \&
  {Scoville}}]{burnham21a}
{Burnham}, A.~D., {Casey}, C.~M., {Zavala}, J.~A., {et~al.} 2021, \apj, 910,
  89, \dodoi{10.3847/1538-4357/abe401}

\bibitem[{{Carnall} {et~al.}(2018){Carnall}, {McLure}, {Dunlop}, \&
  {Dav{\'e}}}]{carnall18a}
{Carnall}, A.~C., {McLure}, R.~J., {Dunlop}, J.~S., \& {Dav{\'e}}, R. 2018,
  \mnras, 480, 4379, \dodoi{10.1093/mnras/sty2169}

\bibitem[{{Casey} {et~al.}(2014{\natexlab{a}}){Casey}, {Narayanan}, \&
  {Cooray}}]{casey14a}
{Casey}, C., {Narayanan}, D., \& {Cooray}, A. 2014{\natexlab{a}}, \physrep,
  541, 45, \dodoi{10.1016/j.physrep.2014.02.009}

\bibitem[{{Casey} {et~al.}(2014{\natexlab{b}})}]{casey14b}
{Casey}, C., {et~al.} 2014{\natexlab{b}}, ApJ, 796, 95

\bibitem[{{Casey} {et~al.}(2017){Casey}, {Cooray}, {Killi}, {Capak}, {Chen},
  {Hung}, {Kartaltepe}, {Sanders}, \& {Scoville}}]{casey17a}
{Casey}, C., {Cooray}, A., {Killi}, M., {et~al.} 2017, \apj, 840, 101,
  \dodoi{10.3847/1538-4357/aa6cb1}

\bibitem[{{Casey} {et~al.}(2018){Casey}, {Zavala}, {Spilker}, {da Cunha},
  {Hodge}, {Hung}, {Staguhn}, {Finkelstein}, \& {Drew}}]{casey18a}
{Casey}, C., {Zavala}, J., {Spilker}, J., {et~al.} 2018, \apj, 862, 77,
  \dodoi{10.3847/1538-4357/aac82d}

\bibitem[{{Casey}(2012)}]{casey12a}
{Casey}, C.~M. 2012, \mnras, 425, 3094,
  \dodoi{10.1111/j.1365-2966.2012.21455.x}

\bibitem[{{Casey}(2020)}]{casey20a}
---. 2020, \apj, 900, 68, \dodoi{10.3847/1538-4357/aba528}

\bibitem[{{Casey}(2024)}]{dustlrds}
---. 2024, {Data for Dust in Little Red Dots}

\bibitem[{{Casey} {et~al.}(2023){Casey}, {Kartaltepe}, {Drakos}, {Franco},
  {Harish}, {Paquereau}, {Ilbert}, {Rose}, {Cox}, {Nightingale}, {Robertson},
  {Silverman}, {Koekemoer}, {Massey}, {McCracken}, {Rhodes}, {Akins}, {Allen},
  {Amvrosiadis}, {Arango-Toro}, {Bagley}, {Bongiorno}, {Capak}, {Champagne},
  {Chartab}, {Ch{\'a}vez Ortiz}, {Chworowsky}, {Cooke}, {Cooper}, {Darvish},
  {Ding}, {Faisst}, {Finkelstein}, {Fujimoto}, {Gentile}, {Gillman}, {Gould},
  {Gozaliasl}, {Hayward}, {He}, {Hemmati}, {Hirschmann}, {Jahnke}, {Jin},
  {Khostovan}, {Kokorev}, {Lambrides}, {Laigle}, {Larson}, {Leung}, {Liu},
  {Liaudat}, {Long}, {Magdis}, {Mahler}, {Mainieri}, {Manning}, {Maraston},
  {Martin}, {McCleary}, {McKinney}, {McPartland}, {Mobasher}, {Pattnaik},
  {Renzini}, {Rich}, {Sanders}, {Sattari}, {Scognamiglio}, {Scoville}, {Sheth},
  {Shuntov}, {Sparre}, {Suzuki}, {Talia}, {Toft}, {Trakhtenbrot}, {Urry},
  {Valentino}, {Vanderhoof}, {Vardoulaki}, {Weaver}, {Whitaker}, {Wilkins},
  {Yang}, \& {Zavala}}]{casey23a}
{Casey}, C.~M., {Kartaltepe}, J.~S., {Drakos}, N.~E., {et~al.} 2023, \apj, 954,
  31, \dodoi{10.3847/1538-4357/acc2bc}

\bibitem[{{da Cunha} {et~al.}(2015){da Cunha}, {Walter}, {Smail}, {Swinbank},
  {Simpson}, {Decarli}, {Hodge}, {Weiss}, {van der Werf}, {Bertoldi},
  {Chapman}, {Cox}, {Danielson}, {Dannerbauer}, {Greve}, {Ivison}, {Karim}, \&
  {Thomson}}]{da-cunha15a}
{da Cunha}, E., {Walter}, F., {Smail}, I., {et~al.} 2015, \apj, 806, 110,
  \dodoi{10.1088/0004-637X/806/1/110}

\bibitem[{{Delhaize} {et~al.}(2017){Delhaize}, {Smol{\v c}i{\'c}},
  {Delvecchio}, {Novak}, {Sargent}, {Baran}, {Magnelli}, {Zamorani},
  {Schinnerer}, {Murphy}, {Aravena}, {Berta}, {Bondi}, {Capak}, {Carilli},
  {Ciliegi}, {Civano}, {Ilbert}, {Karim}, {Laigle}, {Le F{\`e}vre}, {Marchesi},
  {McCracken}, {Salvato}, {Seymour}, \& {Tasca}}]{delhaize17a}
{Delhaize}, J., {Smol{\v c}i{\'c}}, V., {Delvecchio}, I., {et~al.} 2017, \aap,
  602, A4, \dodoi{10.1051/0004-6361/201629430}

\bibitem[{{Diplas} \& {Savage}(1994)}]{diplas94a}
{Diplas}, A., \& {Savage}, B.~D. 1994, \apjs, 93, 211, \dodoi{10.1086/192052}

\bibitem[{{Draine} \& {Li}(2007)}]{draine07a}
{Draine}, B., \& {Li}, A. 2007, \apj, 657, 810, \dodoi{10.1086/511055}

\bibitem[{{Draine} {et~al.}(2007){Draine}, {Dale}, {Bendo}, {Gordon}, {Smith},
  {Armus}, {Engelbracht}, {Helou}, {Kennicutt}, {Li}, {Roussel}, {Walter},
  {Calzetti}, {Moustakas}, {Murphy}, {Rieke}, {Bot}, {Hollenbach}, {Sheth}, \&
  {Teplitz}}]{draine07b}
{Draine}, B., {Dale}, D., {Bendo}, G., {et~al.} 2007, \apj, 663, 866,
  \dodoi{10.1086/518306}

\bibitem[{{Draine} {et~al.}(2014){Draine}, {Aniano}, {Krause}, {Groves},
  {Sandstrom}, {Braun}, {Leroy}, {Klaas}, {Linz}, {Rix}, {Schinnerer},
  {Schmiedeke}, \& {Walter}}]{Draine14a}
{Draine}, B.~T., {Aniano}, G., {Krause}, O., {et~al.} 2014, \apj, 780, 172,
  \dodoi{10.1088/0004-637X/780/2/172}

\bibitem[{{Drew} \& {Casey}(2022)}]{drew22a}
{Drew}, P.~M., \& {Casey}, C.~M. 2022, \apj, 930, 142,
  \dodoi{10.3847/1538-4357/ac6270}

\bibitem[{{Driver} {et~al.}(2018){Driver}, {Andrews}, {da Cunha}, {Davies},
  {Lagos}, {Robotham}, {Vinsen}, {Wright}, {Alpaslan}, {Bland-Hawthorn},
  {Bourne}, {Brough}, {Bremer}, {Cluver}, {Colless}, {Conselice}, {Dunne},
  {Eales}, {Gomez}, {Holwerda}, {Hopkins}, {Kafle}, {Kelvin}, {Loveday},
  {Liske}, {Maddox}, {Phillipps}, {Pimbblet}, {Rowlands}, {Sansom}, {Taylor},
  {Wang}, \& {Wilkins}}]{driver18a}
{Driver}, S.~P., {Andrews}, S.~K., {da Cunha}, E., {et~al.} 2018, \mnras, 475,
  2891, \dodoi{10.1093/mnras/stx2728}

\bibitem[{{Dunne} {et~al.}(2011){Dunne}, {Gomez}, {da Cunha}, {Charlot}, {Dye},
  {Eales}, {Maddox}, {Rowlands}, {Smith}, {Auld}, {Baes}, {Bonfield}, {Bourne},
  {Buttiglione}, {Cava}, {Clements}, {Coppin}, {Cooray}, {Dariush}, {de Zotti},
  {Driver}, {Fritz}, {Geach}, {Hopwood}, {Ibar}, {Ivison}, {Jarvis}, {Kelvin},
  {Pascale}, {Pohlen}, {Popescu}, {Rigby}, {Robotham}, {Rodighiero}, {Sansom},
  {Serjeant}, {Temi}, {Thompson}, {Tuffs}, {van der Werf}, \&
  {Vlahakis}}]{dunne11a}
{Dunne}, L., {Gomez}, H.~L., {da Cunha}, E., {et~al.} 2011, \mnras, 417, 1510,
  \dodoi{10.1111/j.1365-2966.2011.19363.x}

\bibitem[{{Eisenstein} {et~al.}(2023{\natexlab{a}}){Eisenstein}, {Willott},
  {Alberts}, {Arribas}, {Bonaventura}, {Bunker}, {Cameron}, {Carniani},
  {Charlot}, {Curtis-Lake}, {D'Eugenio}, {Endsley}, {Ferruit}, {Giardino},
  {Hainline}, {Hausen}, {Jakobsen}, {Johnson}, {Maiolino}, {Rieke}, {Rieke},
  {Rix}, {Robertson}, {Stark}, {Tacchella}, {Williams}, {Willmer}, {Baker},
  {Baum}, {Bhatawdekar}, {Boyett}, {Chen}, {Chevallard}, {Circosta}, {Curti},
  {Danhaive}, {DeCoursey}, {de Graaff}, {Dressler}, {Egami}, {Helton},
  {Hviding}, {Ji}, {Jones}, {Kumari}, {L{\"u}tzgendorf}, {Laseter}, {Looser},
  {Lyu}, {Maseda}, {Nelson}, {Parlanti}, {Perna}, {Pusk{\'a}s}, {Rawle},
  {Rodr{\'\i}guez Del Pino}, {Sandles}, {Saxena}, {Scholtz}, {Sharpe},
  {Shivaei}, {Silcock}, {Simmonds}, {Skarbinski}, {Smit}, {Stone}, {Suess},
  {Sun}, {Tang}, {Topping}, {{\"U}bler}, {Villanueva}, {Wallace}, {Whitler},
  {Witstok}, \& {Woodrum}}]{eisenstein23a}
{Eisenstein}, D.~J., {Willott}, C., {Alberts}, S., {et~al.} 2023{\natexlab{a}},
  arXiv e-prints, arXiv:2306.02465, \dodoi{10.48550/arXiv.2306.02465}

\bibitem[{{Eisenstein} {et~al.}(2023{\natexlab{b}}){Eisenstein}, {Johnson},
  {Robertson}, {Tacchella}, {Hainline}, {Jakobsen}, {Maiolino}, {Bonaventura},
  {Bunker}, {Cameron}, {Cargile}, {Curtis-Lake}, {Hausen}, {Pusk{\'a}s},
  {Rieke}, {Sun}, {Willmer}, {Willott}, {Alberts}, {Arribas}, {Baker}, {Baum},
  {Bhatawdekar}, {Carniani}, {Charlot}, {Chen}, {Chevallard}, {Curti},
  {DeCoursey}, {D'Eugenio}, {de Graaff}, {Egami}, {Helton}, {Ji}, {Jones},
  {Kumari}, {L{\"u}tzgendorf}, {Laseter}, {Looser}, {Lyu}, {Maseda}, {Nelson},
  {Parlanti}, {Rauscher}, {Rawle}, {Rieke}, {Rix}, {Rujopakarn}, {Sandles},
  {Saxena}, {Scholtz}, {Sharpe}, {Shivaei}, {Simmonds}, {Smit}, {Topping},
  {{\"U}bler}, {Venturi}, {Williams}, {Witstok}, \& {Woodrum}}]{eisenstein23b}
{Eisenstein}, D.~J., {Johnson}, B.~D., {Robertson}, B., {et~al.}
  2023{\natexlab{b}}, arXiv e-prints, arXiv:2310.12340,
  \dodoi{10.48550/arXiv.2310.12340}

\bibitem[{{Gaskell} {et~al.}(2004){Gaskell}, {Goosmann}, {Antonucci}, \&
  {Whysong}}]{gaskell04a}
{Gaskell}, C.~M., {Goosmann}, R.~W., {Antonucci}, R. R.~J., \& {Whysong}, D.~H.
  2004, \apj, 616, 147, \dodoi{10.1086/423885}

\bibitem[{{Greene} {et~al.}(2023){Greene}, {Labbe}, {Goulding}, {Furtak},
  {Chemerynska}, {Kokorev}, {Dayal}, {Williams}, {Wang}, {Setton}, {Burgasser},
  {Bezanson}, {Atek}, {Brammer}, {Cutler}, {Feldmann}, {Fujimoto},
  {Glazebrook}, {de Graaff}, {Leja}, {Marchesini}, {Maseda}, {Matthee},
  {Miller}, {Naidu}, {Nanayakkara}, {Oesch}, {Pan}, {Papovich}, {Price}, {van
  Dokkum}, {Weaver}, {Whitaker}, \& {Zitrin}}]{greene23a}
{Greene}, J.~E., {Labbe}, I., {Goulding}, A.~D., {et~al.} 2023, arXiv e-prints,
  arXiv:2309.05714, \dodoi{10.48550/arXiv.2309.05714}

\bibitem[{{G{\"u}ver} \& {{\"O}zel}(2009)}]{guver09a}
{G{\"u}ver}, T., \& {{\"O}zel}, F. 2009, \mnras, 400, 2050,
  \dodoi{10.1111/j.1365-2966.2009.15598.x}

\bibitem[{{Hickox} \& {Alexander}(2018)}]{hickox18a}
{Hickox}, R.~C., \& {Alexander}, D.~M. 2018, \araa, 56, 625,
  \dodoi{10.1146/annurev-astro-081817-051803}

\bibitem[{{Hodge} {et~al.}(2024){Hodge}, {da Cunha}, {Kendrew}, {Li}, {Smail},
  {Westoby}, {Nayak}, {Swinbank}, {Chen}, {Walter}, {van der Werf}, {Cracraft},
  {Battisti}, {Brandt}, {Calistro Rivera}, {Chapman}, {Cox}, {Dannerbauer},
  {Decarli}, {Frias Castillo}, {Greve}, {Knudsen}, {Leslie}, {Menten}, {Rybak},
  {Schinnerer}, {Wardlow}, \& {Weiss}}]{hodge24a}
{Hodge}, J.~A., {da Cunha}, E., {Kendrew}, S., {et~al.} 2024, arXiv e-prints,
  arXiv:2407.15846, \dodoi{10.48550/arXiv.2407.15846}

\bibitem[{{Hopkins} {et~al.}(2010){Hopkins}, {Murray}, {Quataert}, \&
  {Thompson}}]{hopkins10a}
{Hopkins}, P.~F., {Murray}, N., {Quataert}, E., \& {Thompson}, T.~A. 2010,
  \mnras, 401, L19, \dodoi{10.1111/j.1745-3933.2009.00777.x}

\bibitem[{{Howell} {et~al.}(2010){Howell}, {Armus}, {Mazzarella}, {Evans},
  {Surace}, {Sanders}, {Petric}, {Appleton}, {Bothun}, {Bridge}, {Chan},
  {Charmandaris}, {Frayer}, {Haan}, {Inami}, {Kim}, {Lord}, {Madore},
  {Melbourne}, {Schulz}, {U}, {Vavilkin}, {Veilleux}, \& {Xu}}]{howell10a}
{Howell}, J., {Armus}, L., {Mazzarella}, J., {et~al.} 2010, \apj, 715, 572,
  \dodoi{10.1088/0004-637X/715/1/572}

\bibitem[{{Kennicutt} \& {Evans}(2012)}]{kennicutt12a}
{Kennicutt}, R., \& {Evans}, N. 2012, \araa, 50, 531,
  \dodoi{10.1146/annurev-astro-081811-125610}

\bibitem[{{Kocevski} {et~al.}(2023){Kocevski}, {Onoue}, {Inayoshi}, {Trump},
  {Arrabal Haro}, {Grazian}, {Dickinson}, {Finkelstein}, {Kartaltepe},
  {Hirschmann}, {Fujimoto}, {Juneau}, {Amorin}, {Bagley}, {Barro}, {Bell},
  {Bisigello}, {Calabro}, {Cleri}, {Cooper}, {Ding}, {Grogin}, {Ho}, {Inoue},
  {Jiang}, {Jones}, {Koekemoer}, {Li}, {Li}, {McGrath}, {Molina}, {Papovich},
  {Perez-Gonzalez}, {Pirzkal}, {Wilkins}, {Yang}, \& {Yung}}]{kocevski23a}
{Kocevski}, D.~D., {Onoue}, M., {Inayoshi}, K., {et~al.} 2023, arXiv e-prints,
  arXiv:2302.00012, \dodoi{10.48550/arXiv.2302.00012}

\bibitem[{{Kocevski} {et~al.}(2024){Kocevski}, {Finkelstein}, {Barro},
  {Taylor}, {Calabr{\`o}}, {Laloux}, {Buchner}, {Trump}, {Leung}, {Yang},
  {Dickinson}, {P{\'e}rez-Gonz{\'a}lez}, {Pacucci}, {Inayoshi}, {Somerville},
  {McGrath}, {Akins}, {Bagley}, {Bisigello}, {Bowler}, {Carnall}, {Casey},
  {Cheng}, {Cleri}, {Costantin}, {Cullen}, {Davis}, {Donnan}, {Dunlop},
  {Ellis}, {Ferguson}, {Fujimoto}, {Fontana}, {Giavalisco}, {Grazian},
  {Grogin}, {Hathi}, {Hirschmann}, {Huertas-Company}, {Holwerda},
  {Illingworth}, {Juneau}, {Kartaltepe}, {Koekemoer}, {Li}, {Lucas}, {Magee},
  {Mason}, {McLeod}, {McLure}, {Napolitano}, {Papovich}, {Pirzkal},
  {Rodighiero}, {Santini}, {Wilkins}, \& {Yung}}]{kocevski24a}
{Kocevski}, D.~D., {Finkelstein}, S.~L., {Barro}, G., {et~al.} 2024, arXiv
  e-prints, arXiv:2404.03576, \dodoi{10.48550/arXiv.2404.03576}

\bibitem[{{Kokorev} {et~al.}(2024){Kokorev}, {Caputi}, {Greene}, {Dayal},
  {Trebitsch}, {Cutler}, {Fujimoto}, {Miller}, {Iani}, {Navarro-Carrera}, \&
  {Rinaldi}}]{kokorev24a}
{Kokorev}, V., {Caputi}, K.~I., {Greene}, J.~E., {et~al.} 2024, arXiv e-prints,
  arXiv:2401.09981.
\newblock \doarXiv{2401.09981}

\bibitem[{{Kreckel} {et~al.}(2013){Kreckel}, {Groves}, {Schinnerer}, {Johnson},
  {Aniano}, {Calzetti}, {Croxall}, {Draine}, {Gordon}, {Crocker}, {Dale},
  {Hunt}, {Kennicutt}, {Meidt}, {Smith}, \& {Tabatabaei}}]{kreckel13a}
{Kreckel}, K., {Groves}, B., {Schinnerer}, E., {et~al.} 2013, \apj, 771, 62,
  \dodoi{10.1088/0004-637X/771/1/62}

\bibitem[{{Labbe} {et~al.}(2022){Labbe}, {van Dokkum}, {Nelson}, {Bezanson},
  {Suess}, {Leja}, {Brammer}, {Whitaker}, {Mathews}, \& {Stefanon}}]{labbe23a}
{Labbe}, I., {van Dokkum}, P., {Nelson}, E., {et~al.} 2022, arXiv e-prints,
  arXiv:2207.12446.
\newblock \doarXiv{2207.12446}

\bibitem[{{Labbe} {et~al.}(2023){Labbe}, {Greene}, {Bezanson}, {Fujimoto},
  {Furtak}, {Goulding}, {Matthee}, {Naidu}, {Oesch}, {Atek}, {Brammer},
  {Chemerynska}, {Coe}, {Cutler}, {Dayal}, {Feldmann}, {Franx}, {Glazebrook},
  {Leja}, {Marchesini}, {Maseda}, {Nanayakkara}, {Nelson}, {Pan}, {Papovich},
  {Price}, {Suess}, {Wang}, {Whitaker}, {Williams}, \& {Zitrin}}]{labbe23b}
{Labbe}, I., {Greene}, J.~E., {Bezanson}, R., {et~al.} 2023, arXiv e-prints,
  arXiv:2306.07320, \dodoi{10.48550/arXiv.2306.07320}

\bibitem[{{Li} {et~al.}(2024){Li}, {Inayoshi}, {Chen}, {Ichikawa}, \&
  {Ho}}]{li24a}
{Li}, Z., {Inayoshi}, K., {Chen}, K., {Ichikawa}, K., \& {Ho}, L.~C. 2024,
  arXiv e-prints, arXiv:2407.10760, \dodoi{10.48550/arXiv.2407.10760}

\bibitem[{{Maiolino} {et~al.}(2024){Maiolino}, {Risaliti}, {Signorini},
  {Trefoloni}, {Juodzbalis}, {Scholtz}, {Uebler}, {D'Eugenio}, {Carniani},
  {Fabian}, {Ji}, {Mazzolari}, {Bertola}, {Brusa}, {Bunker}, {Charlot},
  {Comastri}, {Cresci}, {DeCoursey}, {Egami}, {Fiore}, {Gilli}, {Perna},
  {Tacchella}, \& {Venturi}}]{maiolino24a}
{Maiolino}, R., {Risaliti}, G., {Signorini}, M., {et~al.} 2024, arXiv e-prints,
  arXiv:2405.00504, \dodoi{10.48550/arXiv.2405.00504}

\bibitem[{{Matthee} {et~al.}(2024){Matthee}, {Naidu}, {Brammer}, {Chisholm},
  {Eilers}, {Goulding}, {Greene}, {Kashino}, {Labbe}, {Lilly}, {Mackenzie},
  {Oesch}, {Weibel}, {Wuyts}, {Xiao}, {Bordoloi}, {Bouwens}, {van Dokkum},
  {Illingworth}, {Kramarenko}, {Maseda}, {Mason}, {Meyer}, {Nelson}, {Reddy},
  {Shivaei}, {Simcoe}, \& {Yue}}]{matthee24a}
{Matthee}, J., {Naidu}, R.~P., {Brammer}, G., {et~al.} 2024, \apj, 963, 129,
  \dodoi{10.3847/1538-4357/ad2345}

\bibitem[{{McKinney} {et~al.}(2024){McKinney}, {Casey}, {Long}, {Cooper},
  {Manning}, {Franco}, {Akin}, {Lambrides}, {Gammon}, {Silva}, {Gentile},
  {Zavala}, {Amvrosiadis}, {Andika}, {Brinch}, {Champagne}, {Chartab},
  {Drakos}, {Faisst}, {Fujimoto}, {Gillman}, {Gozaliasl}, {Greve}, {Harish},
  {Hayward}, {Hirschmann}, {Ilbert}, {Kalita}, {Kartaltepe}, {Koekemoer},
  {Kokorev}, {Liu}, {Magdis}, {McCracken}, {Rhodes}, {Robertson}, {Talia},
  {Valentino}, \& {Vijayan}}]{mckinney24a}
{McKinney}, J., {Casey}, C.~M., {Long}, A.~S., {et~al.} 2024, arXiv e-prints,
  arXiv:2408.08346, \dodoi{10.48550/arXiv.2408.08346}

\bibitem[{{M{\'e}nard} \& {Fukugita}(2012)}]{menard12a}
{M{\'e}nard}, B., \& {Fukugita}, M. 2012, \apj, 754, 116,
  \dodoi{10.1088/0004-637X/754/2/116}

\bibitem[{{M{\'e}nard} {et~al.}(2010){M{\'e}nard}, {Scranton}, {Fukugita}, \&
  {Richards}}]{menard10a}
{M{\'e}nard}, B., {Scranton}, R., {Fukugita}, M., \& {Richards}, G. 2010,
  \mnras, 405, 1025, \dodoi{10.1111/j.1365-2966.2010.16486.x}

\bibitem[{{Misgeld} \& {Hilker}(2011)}]{misgeld11a}
{Misgeld}, I., \& {Hilker}, M. 2011, \mnras, 414, 3699,
  \dodoi{10.1111/j.1365-2966.2011.18669.x}

\bibitem[{{Nenkova} {et~al.}(2008){Nenkova}, {Sirocky}, {Ivezi{\'c}}, \&
  {Elitzur}}]{nenkova08a}
{Nenkova}, M., {Sirocky}, M., {Ivezi{\'c}}, {\v Z}., \& {Elitzur}, M. 2008,
  \apj, 685, 147, \dodoi{10.1086/590482}

\bibitem[{{Oesch} {et~al.}(2023){Oesch}, {Brammer}, {Naidu}, {Bouwens},
  {Chisholm}, {Illingworth}, {Matthee}, {Nelson}, {Qin}, {Reddy}, {Shapley},
  {Shivaei}, {van Dokkum}, {Weibel}, {Whitaker}, {Wuyts}, {Covelo-Paz},
  {Endsley}, {Fudamoto}, {Giovinazzo}, {Herard-Demanche}, {Kerutt},
  {Kramarenko}, {Labbe}, {Leonova}, {Lin}, {Magee}, {Marchesini}, {Maseda},
  {Mason}, {Matharu}, {Meyer}, {Neufeld}, {Prieto Lyon}, {Schaerer}, {Sharma},
  {Shuntov}, {Smit}, {Stefanon}, {Wyithe}, \& {Xiao}}]{oesch23a}
{Oesch}, P.~A., {Brammer}, G., {Naidu}, R.~P., {et~al.} 2023, arXiv e-prints,
  arXiv:2304.02026, \dodoi{10.48550/arXiv.2304.02026}

\bibitem[{{P{\'e}roux} \& {Howk}(2020)}]{peroux20a}
{P{\'e}roux}, C., \& {Howk}, J.~C. 2020, \araa, 58, 363,
  \dodoi{10.1146/annurev-astro-021820-120014}

\bibitem[{{Peterson}(2006)}]{peterson06a}
{Peterson}, B.~M. 2006, in Physics of Active Galactic Nuclei at all Scales, ed.
  D.~{Alloin}, Vol. 693, 77, \dodoi{10.1007/3-540-34621-X_3}

\bibitem[{{Pope} {et~al.}(2008){Pope}, {Chary}, {Alexander}, {Armus},
  {Dickinson}, {Elbaz}, {Frayer}, {Scott}, \& {Teplitz}}]{pope08a}
{Pope}, A., {Chary}, R., {Alexander}, D., {et~al.} 2008, \apj, 675, 1171,
  \dodoi{10.1086/527030}

\bibitem[{{Pozzi} {et~al.}(2020){Pozzi}, {Calura}, {Zamorani}, {Delvecchio},
  {Gruppioni}, \& {Santini}}]{pozzi20a}
{Pozzi}, F., {Calura}, F., {Zamorani}, G., {et~al.} 2020, \mnras, 491, 5073,
  \dodoi{10.1093/mnras/stz2724}

\bibitem[{{R{\'e}my-Ruyer} {et~al.}(2014){R{\'e}my-Ruyer}, {Madden},
  {Galliano}, {Galametz}, {Takeuchi}, {Asano}, {Zhukovska}, {Lebouteiller},
  {Cormier}, {Jones}, {Bocchio}, {Baes}, {Bendo}, {Boquien}, {Boselli},
  {DeLooze}, {Doublier-Pritchard}, {Hughes}, {Karczewski}, \&
  {Spinoglio}}]{remy-ruyer14a}
{R{\'e}my-Ruyer}, A., {Madden}, S., {Galliano}, F., {et~al.} 2014, \aap, 563,
  A31, \dodoi{10.1051/0004-6361/201322803}

\bibitem[{{Salim} \& {Narayanan}(2020)}]{salim20a}
{Salim}, S., \& {Narayanan}, D. 2020, arXiv e-prints, arXiv:2001.03181.
\newblock \doarXiv{2001.03181}

\bibitem[{{Schneider} \& {Maiolino}(2023)}]{schneider23b}
{Schneider}, R., \& {Maiolino}, R. 2023, arXiv e-prints, arXiv:2310.00053,
  \dodoi{10.48550/arXiv.2310.00053}

\bibitem[{{Scoville} {et~al.}(2016){Scoville}, {Sheth}, {Aussel}, {Vanden
  Bout}, {Capak}, {Bongiorno}, {Casey}, {Murchikova}, {Koda},
  {{\'A}lvarez-M{\'a}rquez}, {Lee}, {Laigle}, {McCracken}, {Ilbert}, {Pope},
  {Sanders}, {Chu}, {Toft}, {Ivison}, \& {Manohar}}]{scoville16a}
{Scoville}, N., {Sheth}, K., {Aussel}, H., {et~al.} 2016, \apj, 820, 83,
  \dodoi{10.3847/0004-637X/820/2/83}

\bibitem[{{Shen} {et~al.}(2020){Shen}, {Hopkins}, {Faucher-Gigu{\`e}re},
  {Alexander}, {Richards}, {Ross}, \& {Hickox}}]{shen20a}
{Shen}, X., {Hopkins}, P.~F., {Faucher-Gigu{\`e}re}, C.-A., {et~al.} 2020,
  \mnras, 495, 3252, \dodoi{10.1093/mnras/staa1381}

\bibitem[{{Simpson} {et~al.}(2019){Simpson}, {Smail}, {Swinbank}, {Chapman},
  {Chen}, {Geach}, {Matsuda}, {Wang}, {Wang}, {Yang}, {Ao}, {Asquith},
  {Bourne}, {Coogan}, {Coppin}, {Gullberg}, {Hine}, {Ho}, {Hwang}, {Ivison},
  {Kato}, {Lacaille}, {Lewis}, {Liu}, {Micha{\l}owski}, {Oteo}, {Sawicki},
  {Scholtz}, {Smith}, {Thomson}, \& {Wardlow}}]{simpson19a}
{Simpson}, J., {Smail}, I., {Swinbank}, A., {et~al.} 2019, \apj, 880, 43,
  \dodoi{10.3847/1538-4357/ab23ff}

\bibitem[{{Smol{\v c}i{\'c}} {et~al.}(2017){Smol{\v c}i{\'c}}, {Novak},
  {Bondi}, {Ciliegi}, {Mooley}, {Schinnerer}, {Zamorani}, {Navarrete},
  {Bourke}, {Karim}, {Vardoulaki}, {Leslie}, {Delhaize}, {Carilli}, {Myers},
  {Baran}, {Delvecchio}, {Miettinen}, {Banfield}, {Balokovi{\'c}}, {Bertoldi},
  {Capak}, {Frail}, {Hallinan}, {Hao}, {Herrera Ruiz}, {Horesh}, {Ilbert},
  {Intema}, {Jeli{\'c}}, {Kl{\"o}ckner}, {Krpan}, {Kulkarni}, {McCracken},
  {Laigle}, {Middleberg}, {Murphy}, {Sargent}, {Scoville}, \&
  {Sheth}}]{smolcic17a}
{Smol{\v c}i{\'c}}, V., {Novak}, M., {Bondi}, M., {et~al.} 2017, \aap, 602, A1,
  \dodoi{10.1051/0004-6361/201628704}

\bibitem[{{Swinbank} {et~al.}(2004){Swinbank}, {Smail}, {Chapman}, {Blain},
  {Ivison}, \& {Keel}}]{swinbank04a}
{Swinbank}, A., {Smail}, I., {Chapman}, S., {et~al.} 2004, \apj, 617, 64,
  \dodoi{10.1086/425171}

\bibitem[{{Swinbank} {et~al.}(2010){Swinbank}, {Smail}, {Longmore}, {Harris},
  {Baker}, {De Breuck}, {Richard}, {Edge}, {Ivison}, {Blundell}, {Coppin},
  {Cox}, {Gurwell}, {Hainline}, {Krips}, {Lundgren}, {Neri}, {Siana},
  {Siringo}, {Stark}, {Wilner}, \& {Younger}}]{swinbank10a}
{Swinbank}, A., {Smail}, I., {Longmore}, S., {et~al.} 2010, \nat, 464, 733,
  \dodoi{10.1038/nature08880}

\bibitem[{{Swinbank} {et~al.}(2014){Swinbank}, {Simpson}, {Smail}, {Harrison},
  {Hodge}, {Karim}, {Walter}, {Alexander}, {Brandt}, {de Breuck}, {da Cunha},
  {Chapman}, {Coppin}, {Danielson}, {Dannerbauer}, {Decarli}, {Greve},
  {Ivison}, {Knudsen}, {Lagos}, {Schinnerer}, {Thomson}, {Wardlow}, {Wei{\ss}},
  \& {van der Werf}}]{swinbank14a}
{Swinbank}, A., {Simpson}, J., {Smail}, I., {et~al.} 2014, \mnras, 438, 1267,
  \dodoi{10.1093/mnras/stt2273}

\bibitem[{{Thacker} {et~al.}(2013){Thacker}, {Cooray}, {Smidt}, {De Bernardis},
  {Mitchell-Wynne}, {Amblard}, {Auld}, {Baes}, {Clements}, {Dariush}, {De
  Zotti}, {Dunne}, {Eales}, {Hopwood}, {Hoyos}, {Ibar}, {Jarvis}, {Maddox},
  {Micha{\l}owski}, {Pascale}, {Scott}, {Serjeant}, {Smith}, {Valiante}, \&
  {van der Werf}}]{thacker13a}
{Thacker}, C., {Cooray}, A., {Smidt}, J., {et~al.} 2013, \apj, 768, 58,
  \dodoi{10.1088/0004-637X/768/1/58}

\bibitem[{{Wang} {et~al.}(2024{\natexlab{a}}){Wang}, {Leja}, {de Graaff},
  {Brammer}, {Weibel}, {van Dokkum}, {Baggen}, {Suess}, {Greene}, {Bezanson},
  {Cleri}, {Hirschmann}, {Labbe}, {Matthee}, {McConachie}, {Naidu}, {Nelson},
  {Oesch}, {Setton}, \& {Williams}}]{wang24b}
{Wang}, B., {Leja}, J., {de Graaff}, A., {et~al.} 2024{\natexlab{a}}, arXiv
  e-prints, arXiv:2405.01473.
\newblock \doarXiv{2405.01473}

\bibitem[{{Wang} {et~al.}(2024{\natexlab{b}}){Wang}, {de Graaff}, {Davies},
  {Greene}, {Leja}, {Goulding}, {Williams}, {Brammer}, {Suess}, {Weibel},
  {Bezanson}, {Boogaard}, {Cleri}, {Hirschmann}, {Katz}, {Labbe}, {Maseda},
  {Matthee}, {McConachie}, {Naidu}, {Oesch}, {Rix}, {Setton}, \&
  {Whitaker}}]{wang24a}
{Wang}, B., {de Graaff}, A., {Davies}, R.~L., {et~al.} 2024{\natexlab{b}},
  arXiv e-prints, arXiv:2403.02304, \dodoi{10.48550/arXiv.2403.02304}

\bibitem[{{Whitaker} {et~al.}(2017){Whitaker}, {Pope}, {Cybulski}, {Casey},
  {Popping}, \& {Yun}}]{whitaker17a}
{Whitaker}, K.~E., {Pope}, A., {Cybulski}, R., {et~al.} 2017, \apj, 850, 208,
  \dodoi{10.3847/1538-4357/aa94ce}

\bibitem[{{Williams} {et~al.}(2023{\natexlab{a}}){Williams}, {Alberts}, {Ji},
  {Hainline}, {Lyu}, {Rieke}, {Endsley}, {Suess}, {Johnson}, {Florian},
  {Shivaei}, {Rujopakarn}, {Baker}, {Bhatawdekar}, {Boyett}, {Bunker},
  {Carniani}, {Charlot}, {Curtis-Lake}, {DeCoursey}, {de Graaff}, {Egami},
  {Eisenstein}, {Gibson}, {Hausen}, {Helton}, {Maiolino}, {Maseda}, {Nelson},
  {Perez-Gonzalez}, {Rieke}, {Robertson}, {Sun}, {Tacchella}, {Willmer}, \&
  {Willott}}]{williams23a}
{Williams}, C.~C., {Alberts}, S., {Ji}, Z., {et~al.} 2023{\natexlab{a}}, arXiv
  e-prints, arXiv:2311.07483, \dodoi{10.48550/arXiv.2311.07483}

\bibitem[{{Williams} {et~al.}(2023{\natexlab{b}}){Williams}, {Tacchella},
  {Maseda}, {Robertson}, {Johnson}, {Willott}, {Eisenstein}, {Willmer}, {Ji},
  {Hainline}, {Helton}, {Alberts}, {Baum}, {Bhatawdekar}, {Boyett}, {Bunker},
  {Carniani}, {Charlot}, {Chevallard}, {Curtis-Lake}, {de Graaff}, {Egami},
  {Franx}, {Kumari}, {Maiolino}, {Nelson}, {Rieke}, {Sandles}, {Shivaei},
  {Simmonds}, {Smit}, {Suess}, {Sun}, {{\"U}bler}, \& {Witstok}}]{williams23b}
{Williams}, C.~C., {Tacchella}, S., {Maseda}, M.~V., {et~al.}
  2023{\natexlab{b}}, \apjs, 268, 64, \dodoi{10.3847/1538-4365/acf130}

\bibitem[{{Zavala} {et~al.}(2021){Zavala}, {Casey}, {Manning}, {Aravena},
  {Bethermin}, {Caputi}, {Clements}, {Cunha}, {Drew}, {Finkelstein},
  {Fujimoto}, {Hayward}, {Hodge}, {Kartaltepe}, {Knudsen}, {Koekemoer}, {Long},
  {Magdis}, {Man}, {Popping}, {Sanders}, {Scoville}, {Sheth}, {Staguhn},
  {Toft}, {Treister}, {Vieira}, \& {Yun}}]{zavala21a}
{Zavala}, J.~A., {Casey}, C.~M., {Manning}, S.~M., {et~al.} 2021, \apj, 909,
  165, \dodoi{10.3847/1538-4357/abdb27}

\bibitem[{{Zimmerman} {et~al.}(2024){Zimmerman}, {Narayanan}, {Whitaker}, \&
  {Dav{\`e}}}]{zimmerman24a}
{Zimmerman}, D.~T., {Narayanan}, D., {Whitaker}, K.~E., \& {Dav{\`e}}, R. 2024,
  arXiv e-prints, arXiv:2401.06719, \dodoi{10.48550/arXiv.2401.06719}

\end{thebibliography}
\end{document}